\documentstyle[preprint,eqsecnum,aps,amsfonts,amssymb,psfig,epsfig,prb]{revtex} 

\draft
\begin{document}


\def\reff#1{(\ref{#1})}
\newcommand{\be}{\begin{equation}}
\newcommand{\ee}{\end{equation}}
\newcommand{\<}{\langle}
\renewcommand{\>}{\rangle}

\def\spose#1{\hbox to 0pt{#1\hss}}
\def\ltapprox{\mathrel{\spose{\lower 3pt\hbox{$\mathchar"218$}}
 \raise 2.0pt\hbox{$\mathchar"13C$}}}
\def\gtapprox{\mathrel{\spose{\lower 3pt\hbox{$\mathchar"218$}}
 \raise 2.0pt\hbox{$\mathchar"13E$}}}

\def\bsigma{\mbox{\protect\boldmath $\sigma$}}
\def\bpi{\mbox{\protect\boldmath $\pi$}}
\def\smfrac#1#2{{\textstyle\frac{#1}{#2}}}
\def\smhalf{ {\smfrac{1}{2}} }

\newcommand{\re}{\mathop{\rm Re}\nolimits}
\newcommand{\im}{\mathop{\rm Im}\nolimits}
\newcommand{\tr}{\mathop{\rm tr}\nolimits}
\newcommand{\fr}{\frac}
\newcommand{\diti}{\frac{\mathrm{d}^2t}{(2 \pi)^2}}

\def\Z{{\mathbb Z}}
\def\R{{\mathbb R}}
\def\C{{\mathbb C}}

\title{Corrections to scaling in multicomponent polymer solutions}

\author{ Andrea Pelissetto }
\address{
  Dipartimento di Fisica and INFN -- Sezione di Roma I  \\
  Universit\`a degli Studi di Roma ``La Sapienza'' \\
  Piazzale Moro 2, I-00185 Roma, Italy \\
  e-mail: {\rm Andrea.Pelissetto@roma1.infn.it }}
\author  {Ettore Vicari}
\address{ Dipartimento di Fisica and INFN -- Sezione di Pisa  \\
  Universit\`a degli Studi di Pisa \\
  Largo Pontecorvo 2, I-56127 Pisa, Italy \\
  e-mail: {\rm Ettore.Vicari@df.unipi.it }}

\maketitle
\thispagestyle{empty}   

\begin{abstract}
\bigskip 

We calculate the correction-to-scaling exponent $\omega_T$
that characterizes the approach to the scaling limit in multicomponent
polymer solutions. 
A direct Monte Carlo determination of $\omega_T$ in a system of interacting
self-avoiding walks gives $\omega_T = 0.415 \pm 0.020$.
A field-theory analysis based on five- and six-loop
perturbative series leads to $\omega_T = 0.41 \pm 0.04$. 
We also verify the renormalization-group predictions for the scaling 
behavior close to the ideal-mixing point.

\bigskip 



\end{abstract}

\clearpage

\section{Introduction}

The behavior of dilute or semidilute solutions of long polymers have been
investigated at length by using the renormalization 
group,\cite{deGennes-79,Freed-87,Schaefer-99} which has explained the scaling
behavior observed in these systems and has provided quantitative predictions
that become exact when the degree of polymerization becomes infinite.
Most of the work has been devoted to binary systems, i.e.~to solutions
of one polymer species in a solvent. The method, however, can be extended to
multicomponent polymer systems, i.e. to solutions of several chemically
different polymers. The general theory has been worked out in detail
in Refs.~\CITE{JLB-84,SK-85,SLK-91}. In the good-solvent regime in which
polymers are swollen, the scaling limit does not change. For instance,
the radius of gyration $R_g$ increases as 
$N^\nu$, where\cite{BN-97,bestnu} $\nu\approx 0.5876$ and $N$ is the 
length of the polymer. However, the presence of chemically different polymers
gives rise to new scaling corrections in quantities that are related to the 
polymer-polymer interaction. In the dilute regime one may consider, for 
instance, the second virial coefficient $B_2$ between two polymers of different 
species. Its scaling behavior is\cite{JLB-84,SK-85,SLK-91}
\begin{equation}
B_2 = A (R_{g,1} R_{g,2})^{3/2} \left[
   1 + a (R_{g,1} R_{g,2})^{-\omega_T/2} + \cdots + 
   b (R_{g,1} R_{g,2})^{-\omega/2} + \cdots \right],
\label{B2eq}
\end{equation}
where $R_{g,1}$ and $R_{g,2}$ are the gyration radii of the two polymers, 
$A$, $a$, $b$ are functions of $R_{g,1}/R_{g,2}$, and $\omega_T$, $\omega$
are correction-to-scaling exponents. Eq.~\reff{B2eq} is valid 
for long polymers in the good-solvent regime; 
more precisely, for $N_1,N_2\to \infty$ at fixed
$N_1/N_2$ (or, equivalently, at fixed $R_{g,1}/R_{g,2}$),
where $N_1$ and $N_2$ are the lengths of the two polymers.
The exponent $\omega$ is the one that controls the scaling
corrections in binary systems. The most accurate estimate of the
exponent so far yields \cite{BN-97} $\Delta = \omega \nu = 
0.517\pm 0.007^{+0.010}_{-0.000}$, corresponding to 
$\omega = 0.880 \pm 0.012^{+0.017}_{-0.000}$. The exponent $\omega_T$ is a
new exponent that characterizes multicomponent systems. 
Perturbative calculations\cite{BLJ-87,SLK-91} indicate that $\omega_T$
is quite small, $\omega_T\approx 0.4$. Thus, scaling corrections decrease
very slowly in multicomponent systems and can be quite relevant for the 
values of $N_1$ and $N_2$ that can be attained in practice. Therefore,
the determination of the scaling behavior in multicomponent systems
may require extrapolations in $N_1$ and $N_2$, which, in turn, 
require a precise knowledge of the scaling exponents.

In this paper we improve the previous determinations\cite{BLJ-87,SLK-91}
of $\omega_T$.  First, we extend the perturbative three-loop calculations of 
Ref.~\CITE{SLK-91}. We analyze the five-loop expansion of $\omega_T$ 
in powers of $d = 4 - \epsilon$, $d$ being the space dimension, and
the six-loop expansion of $\omega_T$ in the 
fixed-dimension massive zero-momentum (MZM) scheme. 
Second, we compute $\omega_T$ by numerical simulations. For this purpose we 
consider interacting self-avoiding walks and compute the second virial 
coefficient. A careful analysis of its scaling behavior provides us with a
estimate of $\omega_T$. We also consider the ideal-mixing point where the 
interaction between the two different chemical species vanishes. 
A renormalization-group analysis of the behavior close to this point
was presented in Ref.~\CITE{SLK-91}. An extensive Monte Carlo simulation
allows us to verify the theoretical predictions.

The paper is organized as follows. In Sec.~\ref{sec2} we present our 
renormalization-group calculations. In Sec.~\ref{sec3} we
determine the correction-to-scaling exponent by means of a Monte Carlo 
simulation. In Sec.~\ref{sec4} we discuss the ideal-mixing point 
where the effective interaction between the two 
chemically different species vanishes.
Conclusions are presented in Sec.~\ref{sec5}.

\section{Perturbative determination of $\omega_T$} \label{sec2}

The starting point of the calculation is the Landau-Ginzburg-Wilson Hamiltonian
\cite{SLK-91}
\begin{eqnarray}
\displaystyle
{\cal H}=& \int d^d x \left\{ {1\over 2} \left[{(\partial_\mu \phi_1)}^2
+{(\partial_\mu \phi_2)}^2+ r_1 \phi_1^2+ r_2 \phi_2^2\right]+\right.\nonumber \\
&\left. \frac{1}{4!} [u_0 \,\phi_1^4+ 2 w_0 \,\phi_1^2 \phi_2^2+v_0
\,\phi_2^4]\right\}, 
\label{Hphi4}
\end{eqnarray}
where $\phi_1$ and $\phi_2$ are $n$-component fields. As usual,
the polymer theory is obtained in the limit $n\to 0$. In this specific
case, the fixed-point structure of the theory is particularly simple and 
is explained in detail in Ref.~\CITE{SLK-91}. 
One finds that the $\beta$ functions satisfy the following properties:
$\beta_u(u,v,w) = \overline{\beta}(u)$,
$\beta_v(u,v,w) = \overline{\beta}(v)$,
$\beta_w(g,g,g) = \overline{\beta}(g)$,
where $\overline{\beta}(g)$ is the $\beta$ function in the vector
$O(n=0)$ $\varphi^4$ model and $u,v,w$ are renormalized four-point couplings
normalized so that $u \approx C u_0$, $v \approx C v_0$, $w \approx C w_0$
at tree level. The relevant fixed point is the symmetric one 
$u^* = v^* = w^* = g^*$, where $g^*$ is the zero of $\overline{\beta}(g)$.
The exponent $\omega_T$ is given by
\begin{equation}
\omega_T = \left. {\partial \beta_w\over \partial w} \right|_{u=v=w=g^*}.
\end{equation}
The exponent $\omega_T$ can be computed directly in the 
$O(n=0)$ $\varphi^4$ model. Indeed, as discussed in Ref.~\CITE{SLK-91},
$\omega_T = - y_4$, where $y_4$ is the renormalization-group dimension
of $\phi_1^2 \phi_2^2$ in the symmetric theory with 
$u_0 = v_0 = w_0$. It corresponds to an $O(2n)$ vector theory and, for 
$n\to 0$, one is back to the $O(n=0)$ $\varphi^4$ model. 
Using the results of Ref.~\CITE{CPV-03} one can also show that,
for $n\to 0$, 
$\phi_1^2 \phi_2^2$ is a spin-4 perturbation of the $O(2n)$ model
and thus $y_4$ is the renormalization-group dimension of the 
cubic-symmetric perturbation $\sum_a \varphi_a^4$ 
of the $O(n=0)$ $\varphi^4$ model.
Thus, one can use the perturbative expansions reported in 
Refs.~\CITE{KS-95} and \CITE{CPV-00}.
The $\epsilon$ expansion of $\omega_T$ is 
\be
\omega_T = \frac{1}{2} \epsilon - \frac{19}{64} \epsilon^2
             + 0.777867  \epsilon^3  - 2.65211 \epsilon^4  + 11.0225 \epsilon^5
            + O(\epsilon^6).
\ee
At order $\epsilon^3$ it agrees with that given in Ref.~\CITE{SLK-91}.
In the fixed-dimension MZM scheme we have at six loops
\be 
\omega_T = -1 + \frac{3}{2} g - {185\over 216} g^2 + 0.916668 g^3 -
1.22868 g^4 + 1.97599 g^5 - 3.59753 g^6 + O(g^7),
\ee
where $g$ is the four-point zero-momentum renormalized coupling normalized so that 
$\overline{\beta}(g) = - g + g^2 + O(g^3)$, as used in, e.g., Ref.~\CITE{LZ-77}; 
the fixed point corresponds to\cite{Nickel-91,CCP-98,PV-98,GZ-98,PV-00} 
$g^* = 1.40 \pm 0.02$.
In order to obtain quantitative predictions, the perturbative 
series must be properly resummed. We use here the conformal-mapping
method\cite{LZ-77,Zinn-Justin-book} that takes into account the large-order
behavior of the perturbative series. 

From the standard $\epsilon$ expansion we obtain in three dimensions ($\epsilon=1$)
$\omega_T = 0.42 \pm 0.04$, while in the fixed-dimension MZM scheme 
we find $\omega_T = 0.37 \pm 0.04$. Using the pseudo-$\epsilon$
expansion, Ref.~\CITE{CP-05} obtained $\omega_T = 0.380 \pm 0.018$ in
the MZM scheme.
Though compatible, the MZM result is lower than the $\epsilon$-expansion
one. This phenomenon also occurs for 
other exponents and is probably related to the nonanalyticity of 
the renormalization-group functions at the fixed 
point.\cite{Nickel-91,Sokal-94,CCP-98,PV-98}

A more precise estimate is obtained by considering 
$\zeta \equiv  \omega_T - \omega/2$. The perturbative expansion of $\zeta$
has smaller coefficients than that of $\omega_T$ 
and thus $\zeta$ can be determined more precisely.
Its $\epsilon$ expansion is 
\be
\zeta  = \frac{1}{32} \epsilon^2 - 0.133936 \epsilon^3 +
         0.490572 \epsilon^4 - 2.41405 \epsilon^5 + O(\epsilon^6)\; .
\ee
The term proportional to $\epsilon$ is missing, while the other coefficients
are smaller by a factor of 5-10 approximately. Similar cancellations
occur in the MZM scheme: 
\begin{eqnarray} 
\zeta &=& -\frac{1}{2} + \frac{1}{2} g -
{85\over 432} g^2 + 0.136823 g^3 \nonumber \\
&& - 0.110394 g^4 + 0.074425 g^5 + 0.024718 g^6 + O(g^7).
\end{eqnarray}
Resumming the perturbative series, we obtain 
\begin{eqnarray}
  \zeta = -0.006 \pm 0.009 && \qquad \hbox{(MZM)},
\nonumber \\
  \zeta = -0.008 \pm 0.012 && \qquad \hbox{($\epsilon$ exp)}.
\end{eqnarray}
We can combine these estimates with those for $\omega$. If we use the 
Monte Carlo result of Ref.~\CITE{BN-97} reported in the 
introduction, we obtain 
\be
\omega_T = 0.433\pm 0.016^{+0.008}_{-0.000}.
\ee
If instead we use the field-theory estimates of $\omega$ reported in 
Ref.~\CITE{GZ-98}, we obtain
\begin{eqnarray}
  \omega_T = 0.399 \pm 0.018 && \qquad \hbox{(MZM)},
\nonumber \\
  \omega_T = 0.407 \pm 0.022 && \qquad \hbox{($\epsilon$ exp)}.
\end{eqnarray}
Collecting results, we estimate 
\be
\omega_T = 0.41 \pm 0.04,
\ee
where the error should be quite conservative.

\section{Monte Carlo results} \label{sec3}

In order to determine $\omega_T$ numerically, we consider lattice self-avoiding
walks (SAWs) with an attractive 
interaction $-\epsilon$ between nonbonded nearest-neighbor
pairs. If $\beta \equiv \epsilon/k T$ is the reduced inverse temperature,
this model describes a polymer in a good solvent as long as 
$\beta < \beta_\theta$, where $\beta_\theta \approx 0.269$ 
corresponds to the collapse $\theta$ transition.\cite{GH-95,stimebetatheta}
We consider two walks with different
interaction energies $\epsilon_1$ and $\epsilon_2$, i.e. with different
$\beta_1$ and $\beta_2$. We assume $\beta_1, \beta_2 < \beta_\theta$, so that 
both walks are in the good-solvent regime. Then, we consider the 
second virial coefficient
\be
B_2(N_1,N_2;\beta_1,\beta_2,\beta_{12}) \equiv  {1\over2} \int d^3{\mathbf r}\,
     \< 1 - e^{-H(1,2)}\>_{{\mathbf 0},{\mathbf r}},
\label{B2def}
\ee
where the statistical average is over all pairs of SAWs such that
the first one starts at the origin, has $N_1$ steps, and 
corresponds to an inverse reduced temperature $\beta_1$; 
the second one starts at  $\mathbf{r}$, has $N_2$ steps, and
corresponds to an inverse reduced temperature $\beta_2$.
Here $H(1,2)$ is the reduced interaction energy: 
$H(1,2) = + \infty$ if the two walks
intersect each other;  otherwise, $H(1,2) = - \beta_{12} {\cal N}_{nnc}$, where
${\cal N}_{nnc}$ is the number of nearest-neighbor contacts between the 
two walks and $\beta_{12} \equiv \epsilon_{12}/k T$ is the reduced
inverse temperature. In order to generate the walks we use the 
pivot algorithm\cite{Lal,MacDonald,Madras-Sokal,Sokal-95b,Kennedy-02}
with a Metropolis test, while the second virial coefficient
is determined by using the hit-or-miss algorithm discussed in
Ref.~\CITE{LMS-95}. 
We study the invariant ratio 
\be
A_2(N_1,N_2;\beta_1,\beta_2,\beta_{12}) = 
  {B_2(N_1,N_2;\beta_1,\beta_2,\beta_{12}) \over 
   [R_g(N_1;\beta_1)  R_g(N_2;\beta_2)]^{3/2}}, 
\ee
where $R_g(N;\beta)$ is the radius of gyration. As we have already discussed,
in the limit $N_1,N_2\to\infty$ at $R_g(N_2;\beta_2)/R_g(N_1;\beta_1)$ fixed,
$A_2$ obeys a scaling law of the form\cite{SLK-91}
\be
A_2(N_1,N_2;\beta_1,\beta_2,\beta_{12}) =  
    f\left({R_g(N_2;\beta_2)\over R_g(N_1;\beta_1)}\right),
\label{EqA2}
\ee
where $f(x)$ is universal. This applies for $\beta_1,\beta_2 < \beta_\theta$
and, as we shall see, for $\beta_{12}$ sufficiently small. Note that all the 
dependence on the inverse temperatures is encoded in a function of a 
single variable. Moreover, $f(x)$ is also the scaling
function associated with a polymer solution made of two different types
of polymers that have the same chemical composition 
(hence $\beta_1 = \beta_2 = \beta_{12}$) but different lengths.
In that case $R_g(N_2;\beta_2)/R_g(N_1;\beta_1) = (N_2/N_1)^\nu$,  so that 
Eq.~\reff{EqA2} implies that $A_2(N_1,N_2;\beta,\beta,\beta) = g(N_1/N_2)$,
with $g(x)=f(x^\nu)$ universal. The function $f(x)$ satisfies the condition 
$f(x) = f (1/x)$ and de Gennes' relation\cite{deGennes-80}
$f(x) \sim x^p$, $p={3/4 - 1/(2\nu)}$, for $x\to 0$.

We will be interested here in the corrections to Eq.~\reff{EqA2}. In the 
scaling limit we can write 
\begin{eqnarray}
A_2(N_1,N_2;\beta_1,\beta_2,\beta_{12}) &=  &
   f(\rho) + 
  \sum_{n+m\ge 1} {a_{nm}(\beta_1,\beta_2,\beta_{12})\over 
            x^{n\omega_T+m\omega}} f_{nm}(\rho) + \cdots
\end{eqnarray}
where $\rho \equiv R_g(N_2;\beta_2)/R_g(N_1;\beta_1)$,
$x \equiv [R_g(N_2;\beta_2)R_g(N_1;\beta_1)]^{1/2}$, 
and we have neglected the contributions of the additional
correction-to-scaling operators with renormalization-group
dimensions $-\omega_i$. They give rise to additional
corrections proportional to $x^{-p}$, 
$p = n\omega_T+m\omega + \sum n_i \omega_i$. Little is known about
$\omega_i$, though we expect them to satisfy
$\omega_i > \omega_T$, $\omega_i > \omega$. 
In the following we will assume that 
all such exponents satisfy $\omega_i \gtrsim 3 \omega_T \approx
\omega_T + \omega$.
The scaling functions
$g_{nm}(\rho)$ are universal once a specific normalization has been chosen.
Instead, the coefficients
$a_{nm}$ depend on the model and, in particular, on the 
specific values of the parameters $\beta_1$, $\beta_2$, and $\beta_{12}$.

\begin{figure}
\centerline{\epsfig{file=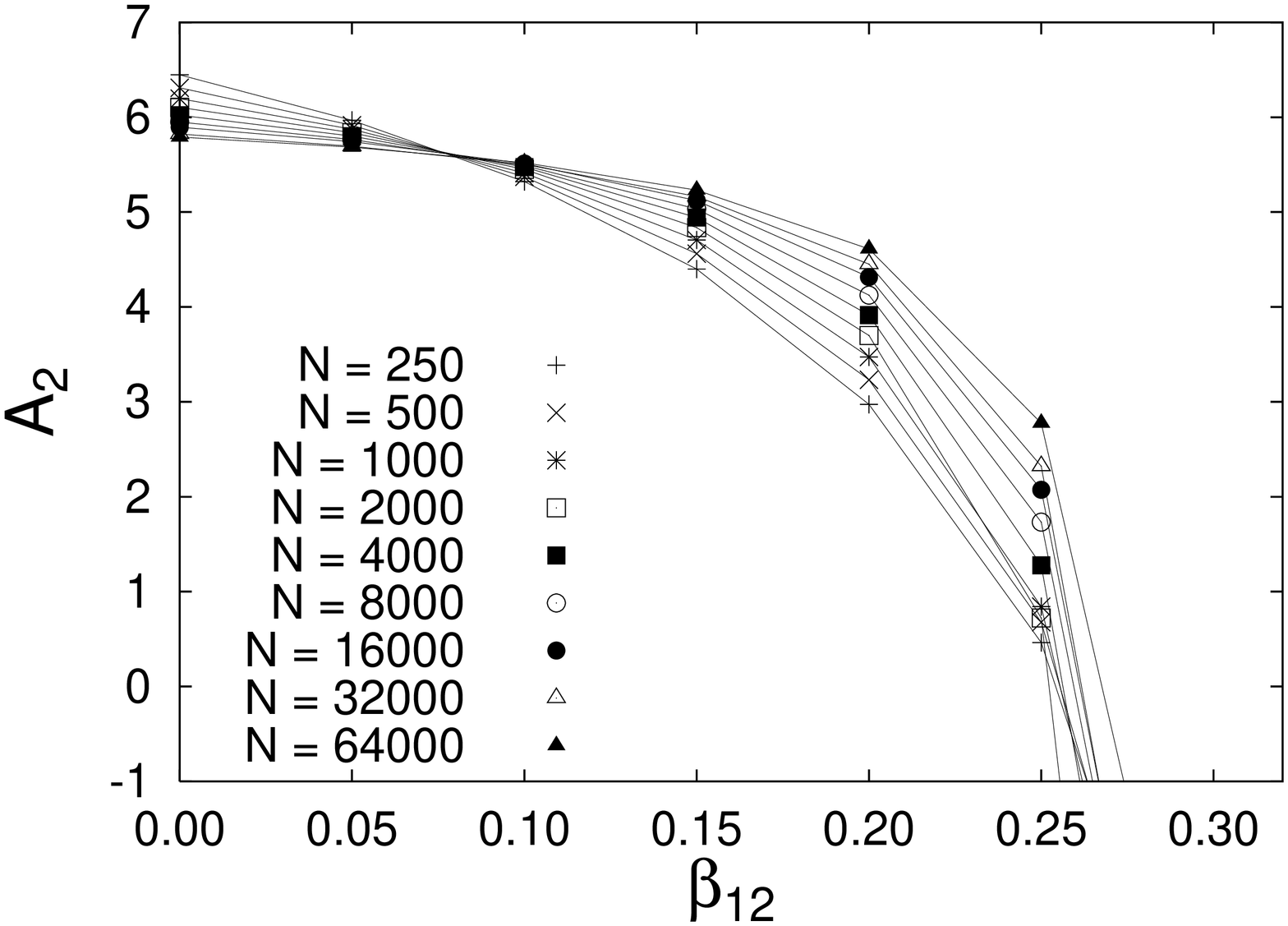,angle=-90,width=10truecm}}
\centerline{\epsfig{file=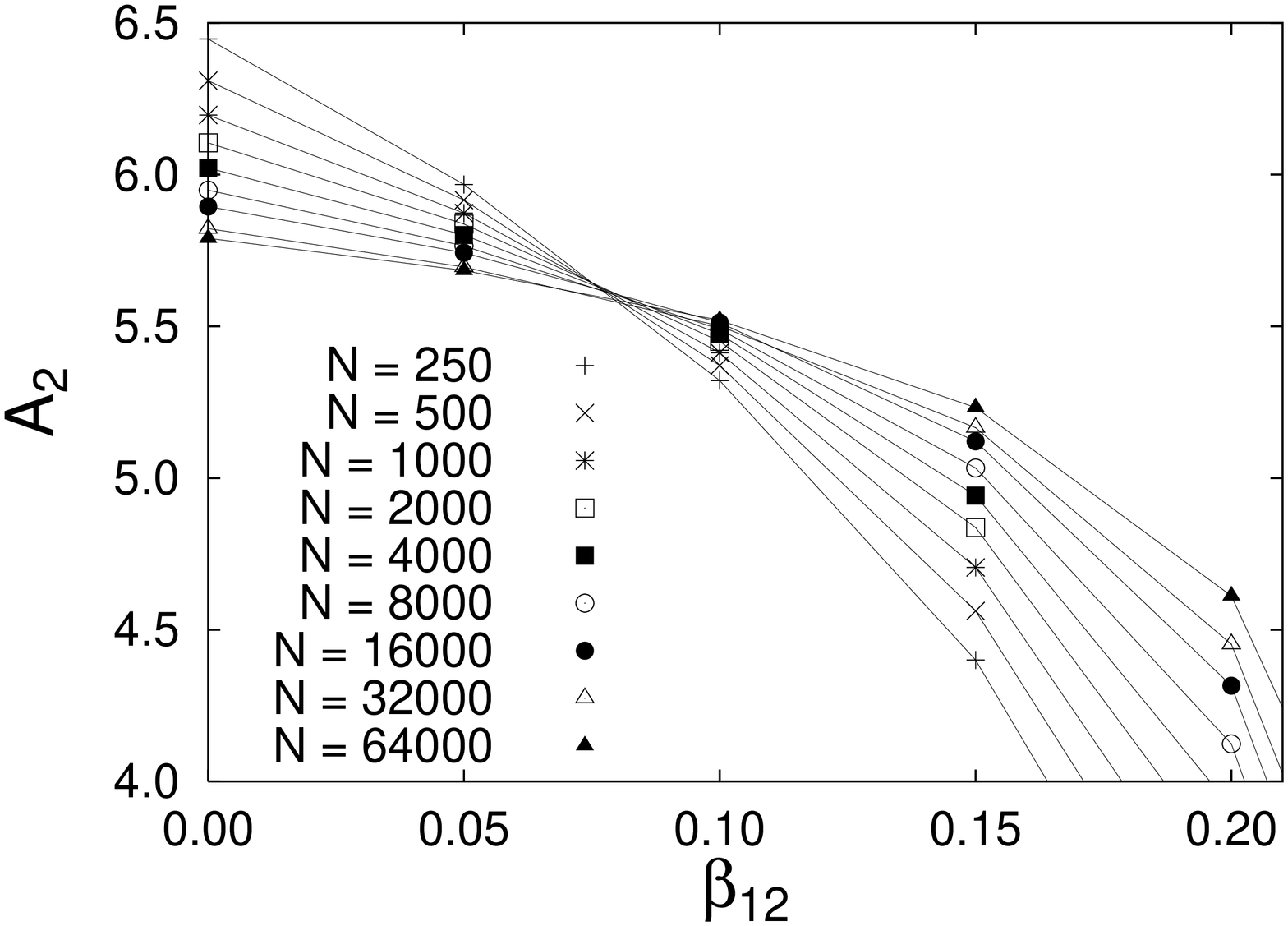,angle=-90,width=10truecm}}
\vspace{1cm}
\caption{Invariant ratio $A_2$ for $\beta_1 = 0.05$, $\beta_2 = 0.15$
vs $\beta_{12}$ for several $N = N_1 = N_2$. The two figures differ only
by the vertical and horizontal scales. Lines connecting points with different
$\beta_{12}$ are only intended to guide the eye.}
\label{Fig1}
\end{figure}

We have simulated two SAWs with $\beta_1 = 0.05$ and $\beta_2 = 0.15$,
two values that are well within the good-solvent region. Then, we have 
computed $A_2$ for $100 \le N_1 = N_2 \le 64000$ and several values of 
$\beta_{12}$ in the range $0\le \beta_{12} \le 0.30$. The results are plotted 
in Fig.~\ref{Fig1}. For $\beta_{12} < 0.25$, as $N=N_1=N_2$ increases, 
the estimates of $A_2$ tend to become independent of $\beta_{12}$ although 
the convergence is very slow. The behavior changes for $\beta_{12}\gtrsim 0.25$
and indeed the data indicate that $A_2 = 0$ for $N\to \infty$ for some 
$\beta_{12} = \beta_{12,c}$ slightly larger than 0.25. This value 
corresponds to the case in which the short-distance repulsion is 
exactly balanced by the solvent-induced attraction proportional to 
$\beta_{12}$. In field-theoretical terms, this means that the 
renormalization-group flow is no longer attracted by the symmetric fixed point
discussed in Sec.~\ref{sec2}, but rather by the unstable 
fixed point with $w^* = 0$.
Thus, at $\beta_{12} = \beta_{12,c}$ there is effectively 
no interaction between the 
chemically different polymers. 
For $\beta_{12} > \beta_{12,c}$,
$A_2$ becomes negative signalling demixing.
The behavior of $A_2$ does not change significantly if $\beta_1$ and $\beta_2$ 
are varied.
In Fig.~\ref{Fig2} we report results for shorter walks for different 
pairs of $\beta_1$ and $\beta_2$, and also for walks with $N_1 = 4 N_2$.
Note that $\beta_{12,c}$ depends very little on the parameters 
$\beta_1$ and $\beta_2$.

In order to determine $\omega_T$, we have considered the data with 
$\beta_{12} = 0, 0.05, 0.10, 0.15$ that are sufficiently far from the 
critical value $\beta_{12,c}$. 
We performed a fit that is linear in $\omega_T$ 
\be
\ln [A_2(N;\beta_{12}=0) - A_2(N;\beta_{12})] = 
c_1(\beta_{12}) - \omega_T \ln x + 
{c_2(\beta_{12})\over x^{\omega_a}} + 
{c_3(\beta_{12})\over x^{\omega_b}}  ,
\ee
where $\beta_{12} = 0.05, 0.10, 0.15$. The exponents $\omega_a$ and 
$\omega_b$ should take into account the additional scaling corrections. Since
$2 \omega_T \approx \omega$, we should have 
$\omega_a \approx \omega_T\approx \omega - \omega_T$. 
Using the field-theory estimate of $\omega_T$ 
and $\omega \approx 0.88 \pm 0.03$ (Ref.~\CITE{BN-97}),
it should be safe to take\cite{footnote} $\omega_a = 0.43 \pm 0.06$.
As for $\omega_b$ we should have 
$\omega_b \approx 2 \omega_T\approx \omega$. Moreover, there is also 
the possibility that there exists an additional correction exponent 
$\omega_1$ not very much different from  $3 \omega_T$, which would 
contribute a correction with exponent $\omega_1 - \omega_T$. 
For this reason we have taken 
$\omega_b = 0.85 \pm 0.20$. The error should be large enough to include all 
possibilities.
The results are reported in Table~\ref{Table}. The systematic error reported
there gives the variation of the estimate as $\omega_a$ and $\omega_b$ vary 
within the reported errors. The results with $N_{\rm min} = 250$ and 500
are compatible within errors and thus we can take the estimate 
that corresponds to $N_{\rm min} = 500$ as our final result.
To be conservative, however, the error bar takes also 
into account the possibility that the observed small trend is a real
one. If the neglected corrections are of order $x^{-3\omega_T}$ 
we expect the results to depend on $N_{\rm min}$ as 
$N_{\rm min}^{-3\nu \omega_T} \approx N_{\rm min}^{-0.7}$. This implies that 
the estimate of $\omega_T$ can decrease at most by 0.005 when 
$N_{\rm min}$ is further decreased. This leads to the 
result
\be
\omega_T = 0.415 \pm 0.020.
\ee
As a check we perform a nonlinear fit of the form (fit 2)
\be
A_2(N;\beta_{12}) = A_2^* +  {a_1(\beta_{12})\over x^{\omega_T}} + 
      {a_2(\beta_{12})\over x^{2\omega_T}} + 
      {a_3(\beta_{12})\over x^{3 \omega_T}};
\label{fit3}
\ee
since, $x \sim N^\nu$, we can also fit the data to (fit 3)
\be
A_2(N;\beta_{12}) = A_2^* +  {a_1(\beta_{12})\over N^{\Delta_T}} + 
      {a_2(\beta_{12})\over N^{2\Delta_T}} + 
      {a_3(\beta_{12})\over N^{3 \Delta_T}},
\label{fit4}
\ee
where $\Delta_T = \omega_T \nu$.
The results are reported in Table~\ref{Table}. They agree with those obtained 
before and allow us to estimate the universal constant $A_2^*$ 
[$A_2^* = f(\rho)$ for $\rho \approx 1.24$; $f(\rho)$ is defined
in Eq.~\reff{EqA2}]: $A_2^* =5.495 \pm 0.020$.

\section{Ideal-mixing point} \label{sec4}

In this section we consider the behavior close to the ideal-mixing point (IMP)
$\beta_{12,c}$ where the effective interaction between the two 
chemically different species vanishes. The renormalization-group 
analysis is presented in Ref.~\CITE{SLK-91}. The behavior is controlled
by an unstable fixed point characterized by an unstable direction
with renormalization-group dimension $y_I = 2/\nu - 3$ and by a stable 
direction with exponent $-\omega$, where $\omega\approx 0.88$ is the usual 
correction-to-scaling exponent. These results imply that close to the 
IMP a renormalization-group invariant quantity ${\cal R}$ (for instance, the 
invariant ratio $A_2$ introduced above) scales as 
\begin{eqnarray}
{\cal R}(\beta_{12}) &= &
   f_{\cal R}[(\beta_{12} - \beta_{12,c}) (R_{g,1} R_{g,2})^{y_I/2}, 
    R_{g,1}/R_{g,2}] 
\nonumber \\ 
  && + {1\over (R_{g,1} R_{g,2})^{\omega/2}}
   g_{\cal R}[(\beta_{12} - \beta_{12,c}) (R_{g,1} R_{g,2})^{y_I/2},
    R_{g,1}/R_{g,2}].
\end{eqnarray}
In this section we wish to verify this scaling behavior for the 
second virial coefficient. For this purpose we have made simulations
for six different pairs of $\beta_1$ and $\beta_2$ with $N_1 = N_2$
and $0.25\le \beta_{12}\le 0.28$. Since $N_1 = N_2 = N$ and 
$R_g(N;\beta) \approx a(\beta) N^\nu$, we can rewrite the previous equation as 
\begin{eqnarray}
&& {\cal R}(N;\beta_1,\beta_2,\beta_{12}) = 
   \hat{f}_{\cal R}(b, \rho) + 
  {1\over N^{\Delta}} \hat{g}_{\cal R}(b,\rho), 
\nonumber 
\\
&& b\equiv (\beta_{12} - \beta_{12,c}) N^\phi, 
\nonumber \\
&& \rho \equiv R_g(N;\beta_1)/R_g(N;\beta_2),
\label{scal1}
\end{eqnarray}
where $\beta_{12,c}$ also depends on $\beta_1$ and $\beta_2$, and
\be
\phi \equiv  \nu y_I = 2 - 3 \nu = 0.2372 \pm 0.0003.
\ee
The critical value $\beta_{12,c}$ can be characterized by requiring 
$A_2(N\to\infty;\beta_1,\beta_2,\beta_{12,c}) = 0$. We can also 
define a finite-$N$ IMP as the value of $\beta_{12}$ where 
$A_2(N;\beta_1,\beta_2,\beta_{12})$ vanishes (this is the analogous
of the Boyle point in $\theta$ solutions):
we define $\beta_{12,c}^{\rm eff}(N)$ such that 
\be
   A_2(N;\beta_1,\beta_2,\beta_{12,c}^{\rm eff}(N)) = 0.
\ee
Inserting in Eq.~(\ref{scal1}) we obtain for $N\to\infty$ the behavior
\be
\beta_{12,c}^{\rm eff}(N) = \beta_{12,c} + {a\over N^{\omega+\phi}}.
\label{beta12eff}
\ee
Finally, we can replace $\beta_{12,c}$ with $\beta_{12,c}^{\rm eff}(N)$
in Eq.~(\ref{scal1}) obtaining the equivalent form
\be
{\cal R}(N;\beta_1,\beta_2,\beta_{12}) =
   \hat{f}_{\cal R}[(\beta_{12} - \beta_{12,c}^{\rm eff}) N^\phi, \rho] + 
  {1\over N^{\Delta}}
   \bar{g}_{\cal R}[(\beta_{12} - \beta_{12,c}^{\rm eff}) N^\phi,\rho],
\label{scal2}
\ee
where $\bar{g}_{\cal R}(b,\rho)$ vanishes at the IMP $b = 0$. 
Eq.~(\ref{scal2}) is more suitable for a numerical check close to the 
IMP than Eq.~(\ref{scal1}), since scaling corrections vanish at the IMP
and are therefore small close to it.
In the following we verify numerically predictions
(\ref{beta12eff}) and (\ref{scal2}).

In Fig.~\ref{Fig:betaeff} we show $\beta_{12,c}^{\rm eff}(N)$ 
vs $N^{-\omega-\phi}$ for three different pairs of $\beta_1$ and $\beta_2$.
The data show a quite good linear behavior: only the two 
points corresponding to $N=2000$ and $4000$ are in some cases 
off the linear fit, probably because our sampling
is not yet adequate for these large values of $N$. 
These results allow us to obtain $\beta_{12,c}$:
\begin{eqnarray}
\beta_{12,c} = 0.2574 \pm 0.0006 \qquad\qquad 
   \hbox{$\beta_1 = 0.05$, $\beta_2 = 0.10$}, \nonumber \\
\beta_{12,c} = 0.2588 \pm 0.0007 \qquad\qquad 
   \hbox{$\beta_1 = 0.05$, $\beta_2 = 0.15$}, \nonumber \\
\beta_{12,c} = 0.2609 \pm 0.0003 \qquad\qquad 
   \hbox{$\beta_1 = 0.05$, $\beta_2 = 0.20$}, \nonumber \\
\beta_{12,c} = 0.2596 \pm 0.0004 \qquad\qquad 
   \hbox{$\beta_1 = 0.10$, $\beta_2 = 0.15$}, \nonumber \\
\beta_{12,c} = 0.2609 \pm 0.0009 \qquad\qquad 
   \hbox{$\beta_1 = 0.10$, $\beta_2 = 0.20$}, \nonumber \\
\beta_{12,c} = 0.2626 \pm 0.0009 \qquad\qquad 
   \hbox{$\beta_1 = 0.15$, $\beta_2 = 0.20$}.
\nonumber 
\end{eqnarray}
Note that the dependence on $\beta_1$ and $\beta_2$ is tiny. 

Then, we verify Eq.~(\ref{scal2}). In Fig.~\ref{Fig:A2scal} we report 
$A_2$ vs the scaling variable $(\beta_{12} - \beta_{12,c}^{\rm eff}) N^\phi$
for different $\beta_1$, $\beta_2$.
All points with $100\le N \le 1000$ fall on top of each other confirming 
the scaling behavior predicted by the renormalization group. 
The scaling functions depend on $\beta_1$ and $\beta_2$ through the 
ratio $\rho$. Moreover, one should also 
take into account that the scaling variable is only defined up to an arbitrary 
prefactor. We now show that the dependence on 
$\rho$ is tiny for the range of values of $\rho$
we have considered, since all data with different values of $N$, $\beta_1$,
and $\beta_2$ fall on a single curve once one takes as scaling
variable $R(\beta_1,\beta_2) (\beta_{12} - \beta_{12,c}^{\rm eff}) N^\phi$,
where $R(\beta_1,\beta_2)$ is a properly chosen constant that 
depends on $\beta_1$ and $\beta_2$. This is evident in Fig.~\ref{Fig:A2resc},
where we report data with different  $\beta_1$ and $\beta_2$. 
We only consider $N=500$ for clarity, since we have already verified that 
data with different values of $N$ show the predicted scaling. 
The scaling is very good, indicating that the $\rho$ dependence is 
negligible.  If we choose $R(0.05,0.10)= 1$ the scaling curve 
can be parametrized as $A_2 = -19.773 x - 46.457 x^2$, with 
$x = R(\beta_1,\beta_2) (\beta_{12} - \beta_{12,c}^{\rm eff}) N^\phi$.
Finally, note that $R(\beta_1,\beta_2)$ does not depend 
very much on $\beta_1$ and $\beta_2$. For instance,
$R(0.15,0.20)/R(0.05,0.10) \approx 0.88$.

\section{Conclusions} \label{sec5}

In this paper we have considered the corrections to scaling expected in 
multicomponent polymer solutions. A high-precision Monte Carlo 
simulation with very long walks, up to $N = 64000$, gives 
$\omega_T = 0.415 \pm 0.020$. This estimate is consistent with 
the estimate $\omega_T = 0.41 \pm 0.04$ obtained by 
using the perturbative renormalization group.
Previous perturbative 
renormalization-group calculations gave $\omega_T \approx 0.37$
(Ref.~\CITE{BLJ-87}) and $\omega_T \approx 0.40$
(Ref.~\CITE{SLK-91}): they substantially agree with our estimate.
On the other hand, the numerical result\cite{SLK-91} $\omega_T \approx 0.35$
obtained by exploiting the relation between
$\omega_T$ and the growth exponent for four-arm star polymers
seems slightly too small.

One should note that $\omega_T$ is quite small and thus convergence may
be quite slow. For instance, in the case we have considered numerically,
$A_2^* = 5.495\pm0.020$ for $N\to \infty$. 
On the other hand, for $\beta_{12} = 0$
(resp. $\beta_{12} = 0.15$) we find $A_2 = 5.790\pm 0.005$ (resp. 
$A_2 = 5.232 \pm 0.005$) for $N = N_1 = N_2 = 64000$. Even if the walks are 
very long, there is still a 5\% discrepancy. For $N = 1000$, 
differences are larger, approximately of 15\%.

\begin{figure}

\begin{minipage}{17pc}
\includegraphics[width=16pc,angle=0]{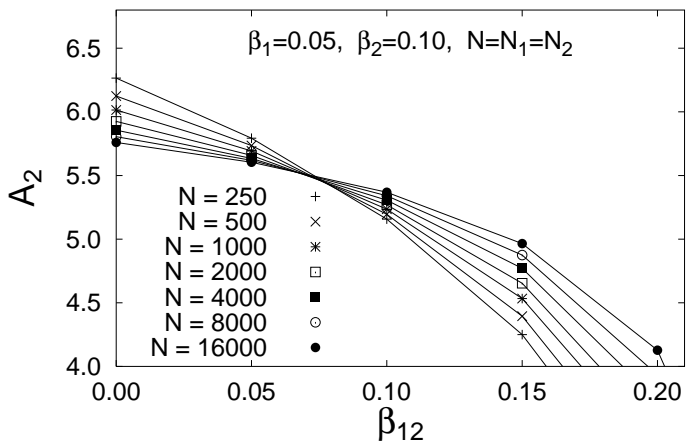}
\end{minipage}\hspace{2pc}%
\begin{minipage}{17pc}
\includegraphics[width=16pc]{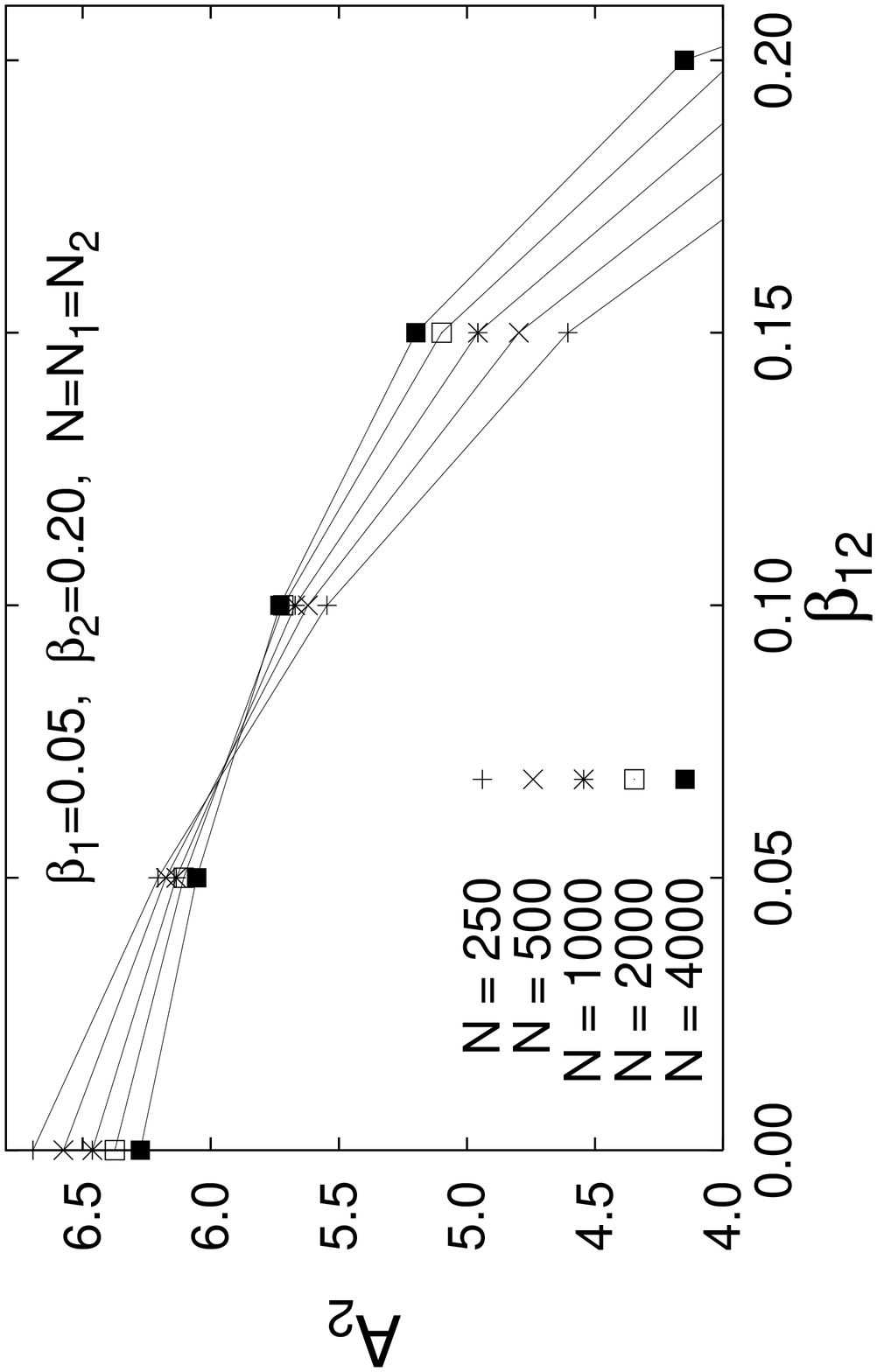}
\end{minipage}

\begin{minipage}{17pc}
\includegraphics[width=16pc]{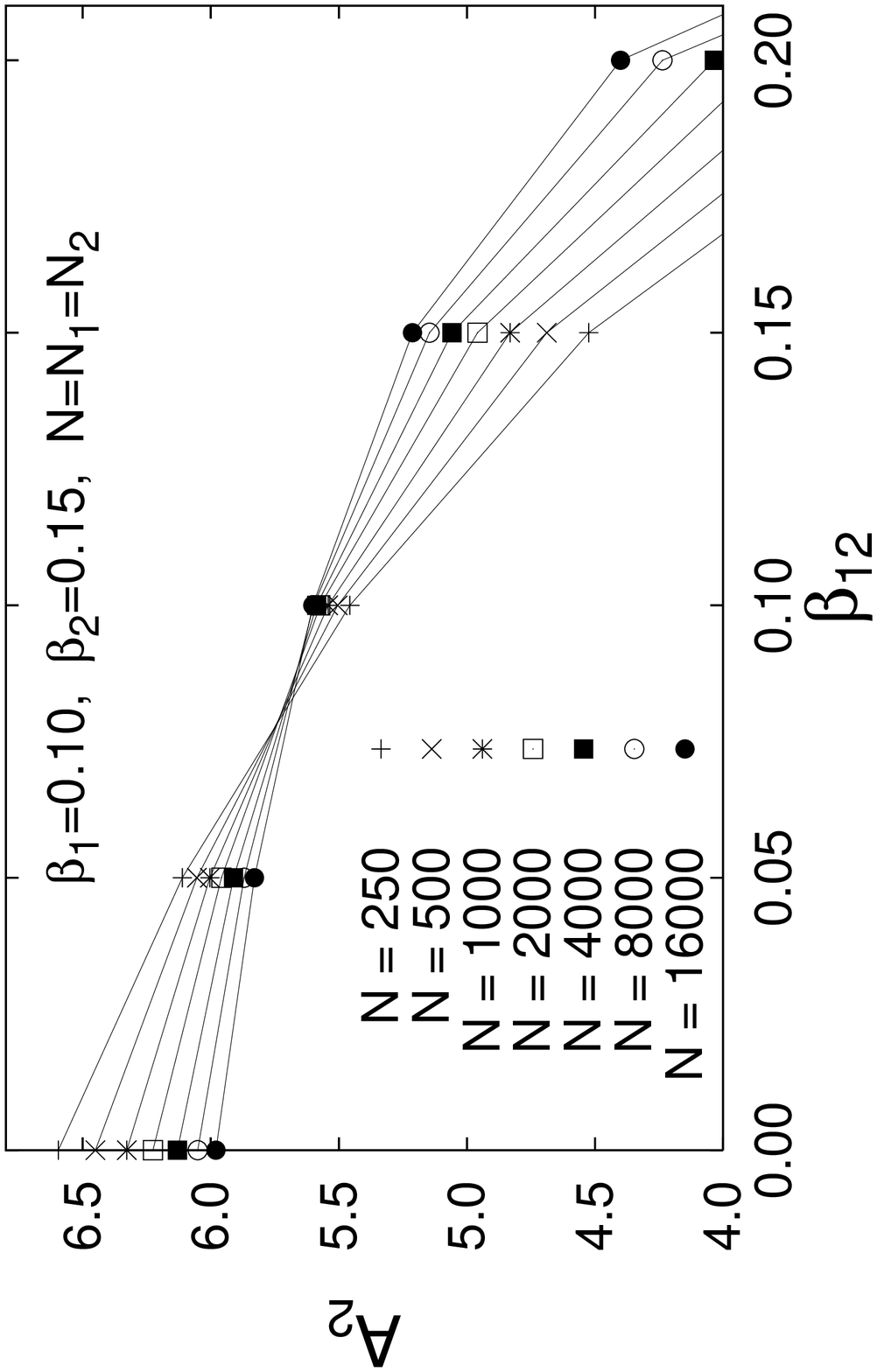}
\end{minipage}\hspace{2pc}%
\begin{minipage}{17pc}
\includegraphics[width=16pc,angle=0]{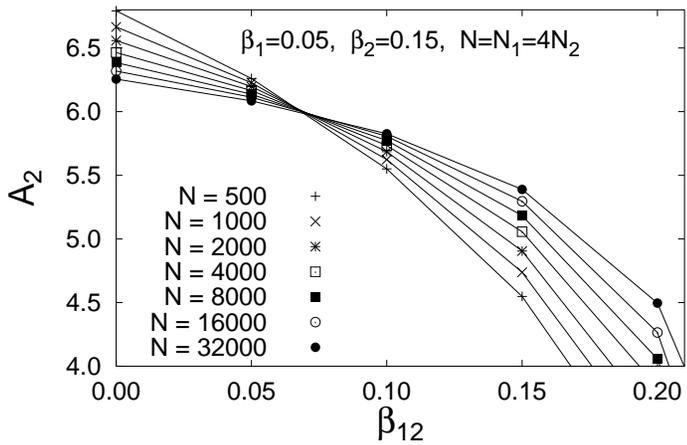}
\end{minipage}

\vspace{1cm}
\caption{Invariant ratio $A_2$ for different choices of $\beta_1$ and 
$\beta_2$ vs $\beta_{12}$.}
\label{Fig2}
\end{figure}

\begin{figure}
\centerline{\epsfig{file=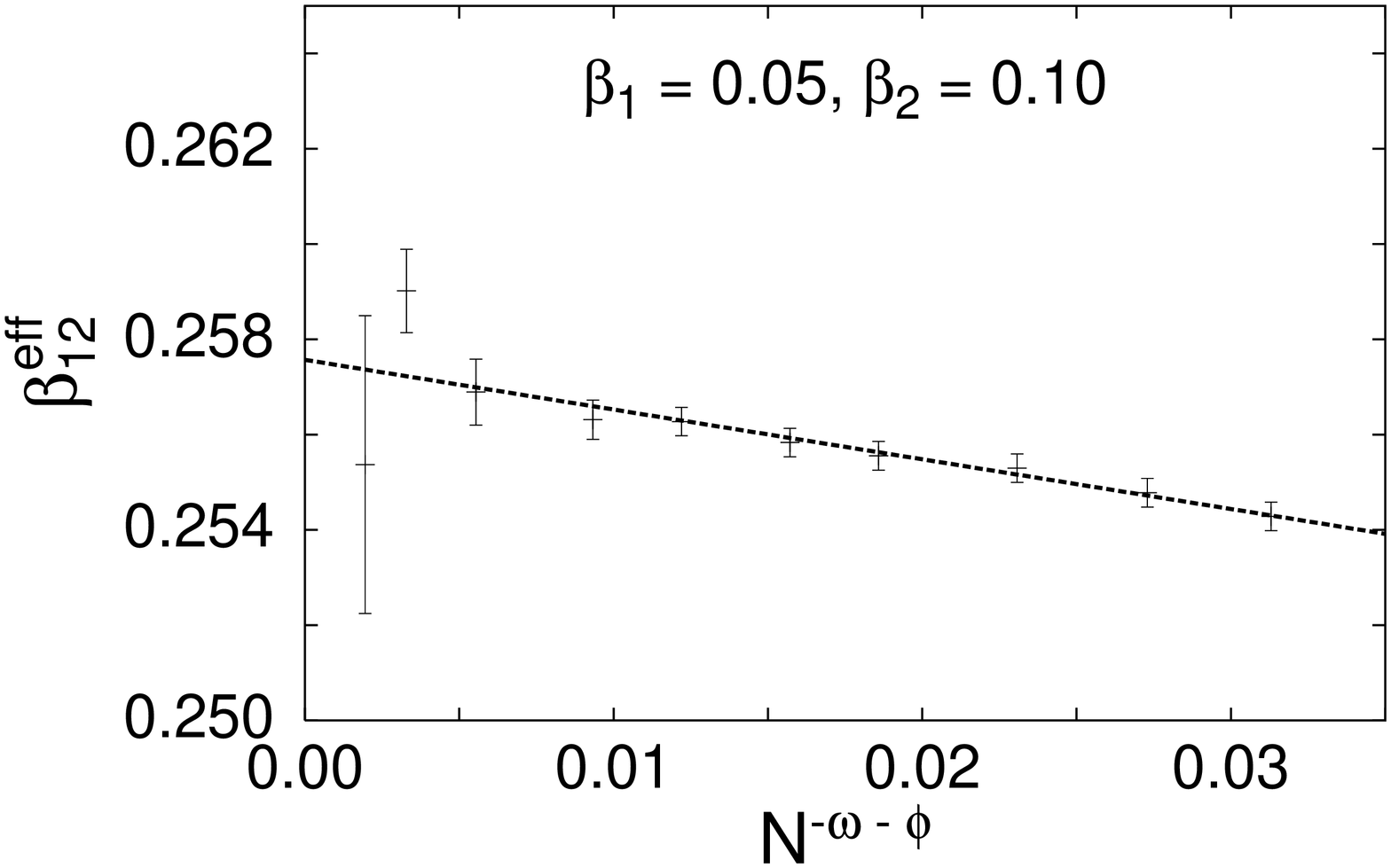,angle=-90,width=9truecm}}
\centerline{\epsfig{file=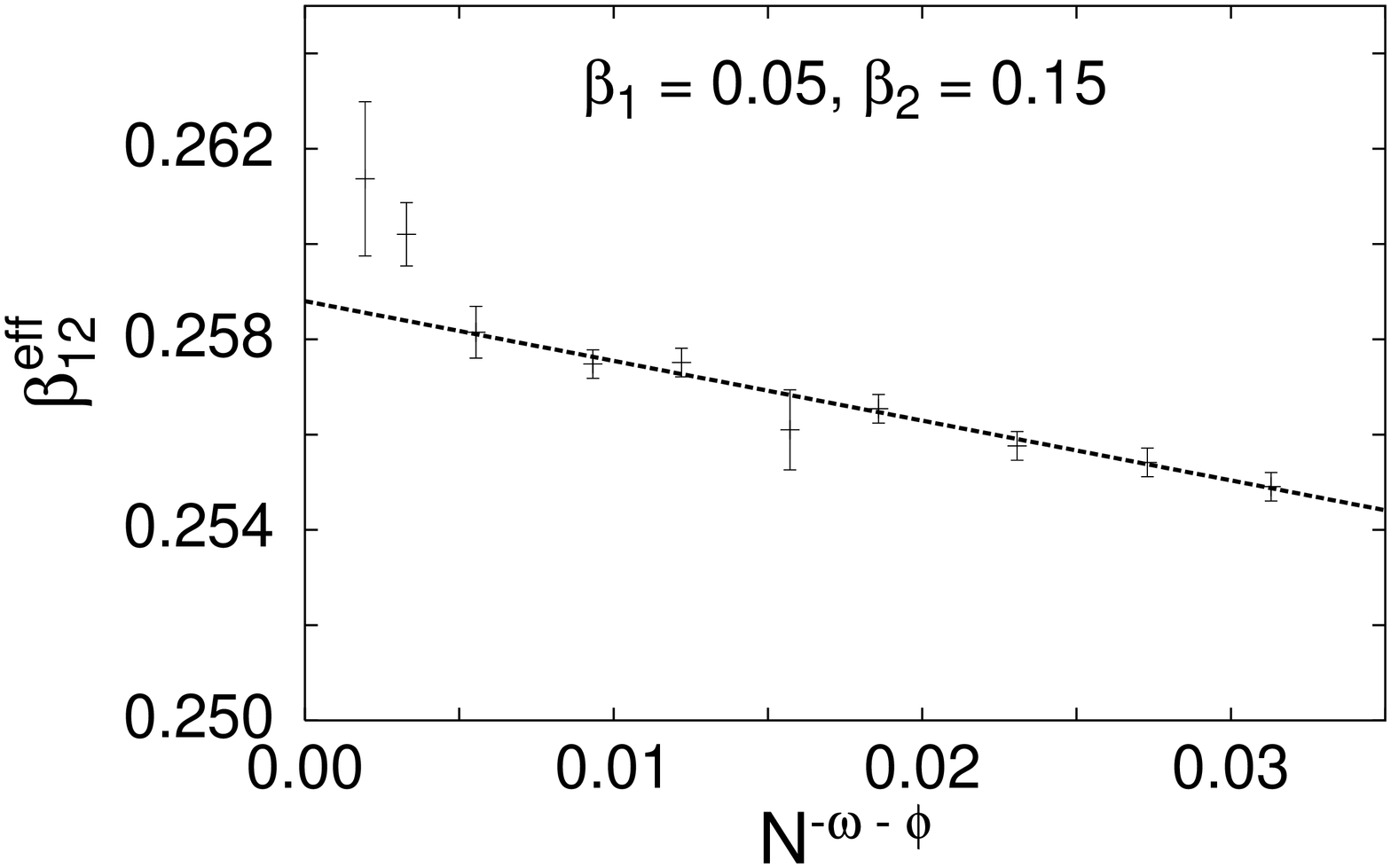,angle=-90,width=9truecm}}
\centerline{\epsfig{file=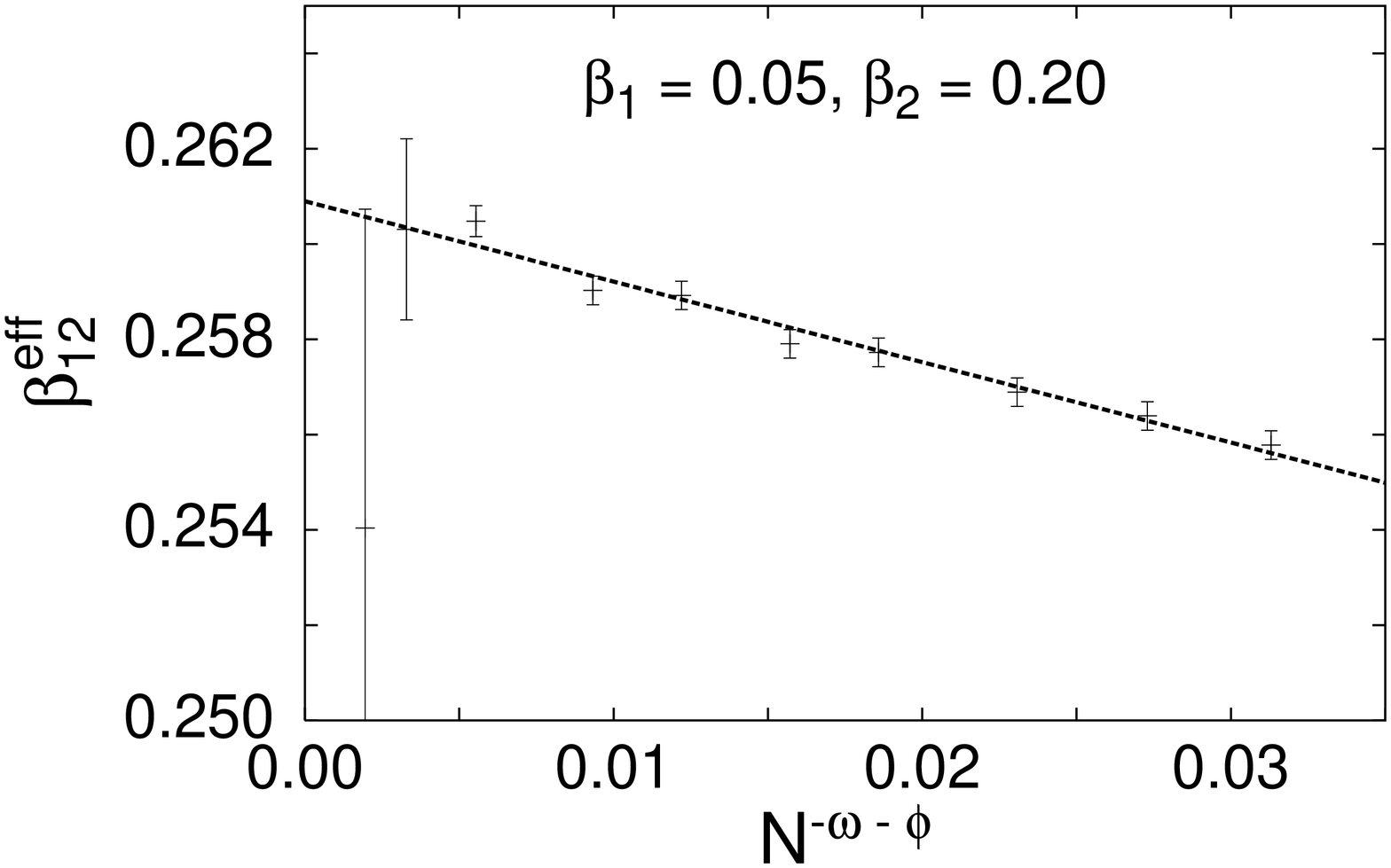,angle=-90,width=9truecm}}
\vspace{1cm}
\caption{Finite-$N$ ideal-mixing point  $\beta_{12,c}^{\rm eff}$
vs $1/N^{\omega+\phi}$ for three different pairs of 
$\beta_1$ and $\beta_2$. The line corresponds to the fit
$\beta_{12,c}^{\rm eff} = \beta_{12,c} + a/N^{\omega+\phi}$.
}
\label{Fig:betaeff}
\end{figure}

\begin{figure}
\centerline{\epsfig{file=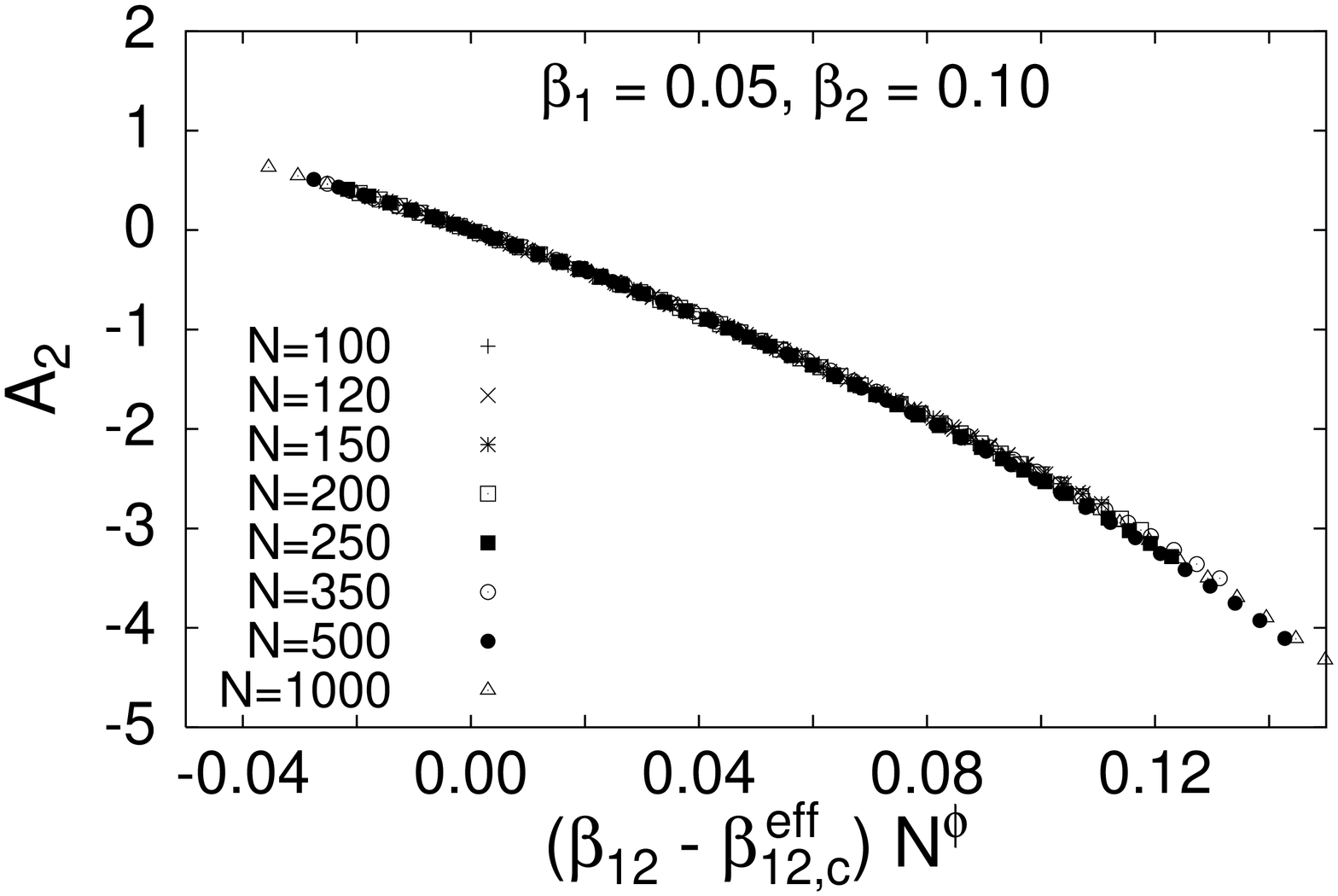,angle=-90,width=9truecm}}
\centerline{\epsfig{file=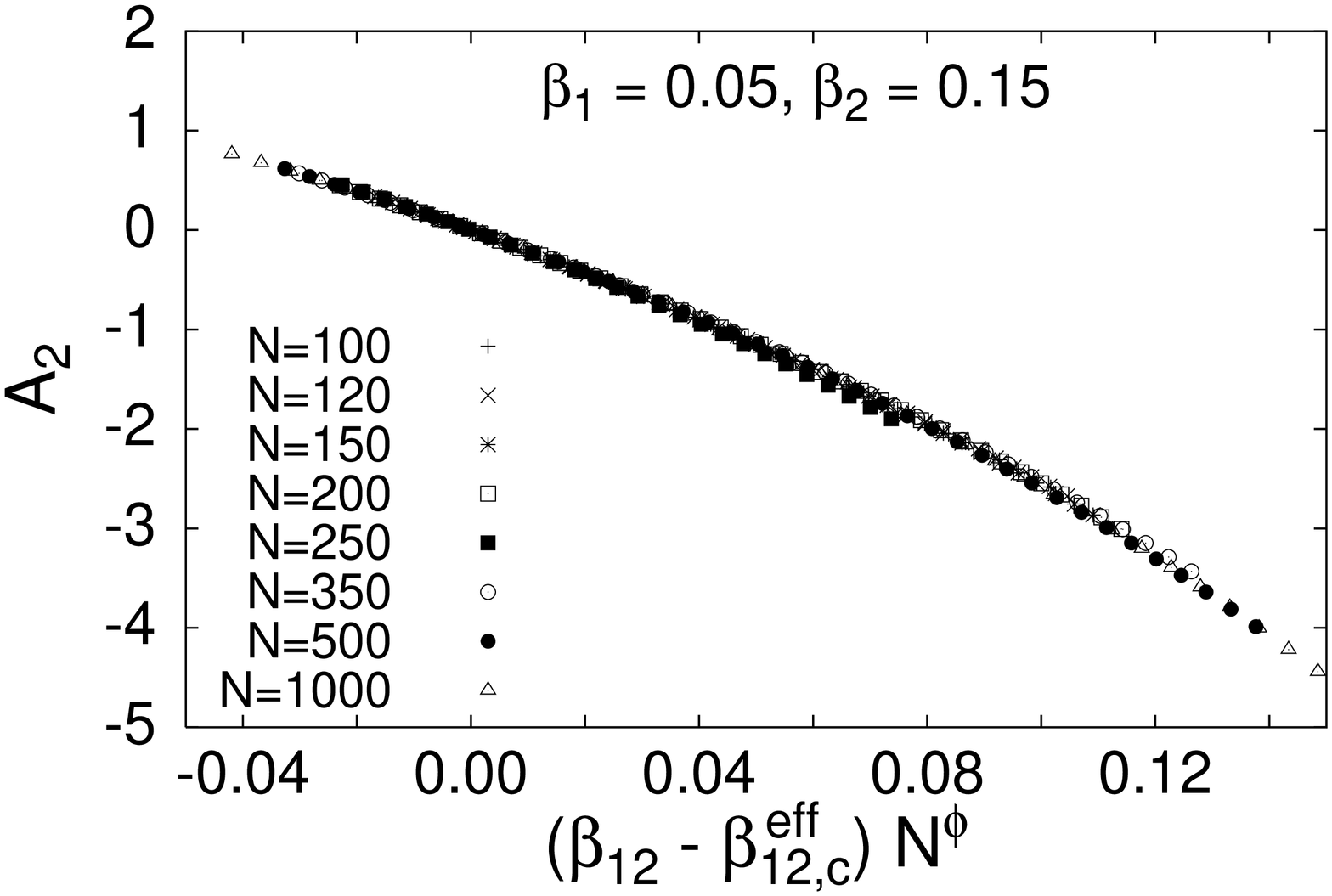,angle=-90,width=9truecm}}
\centerline{\epsfig{file=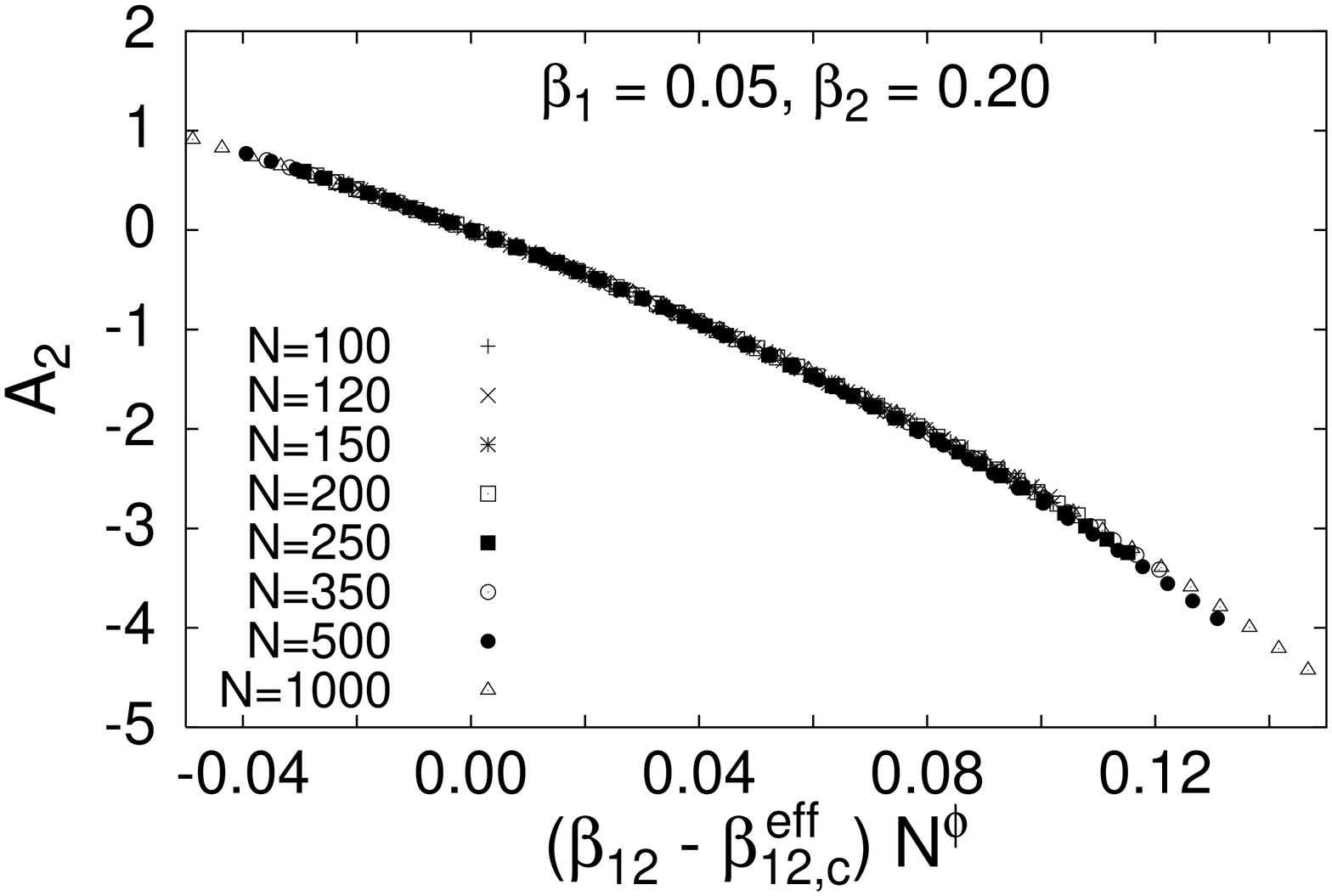,angle=-90,width=9truecm}}
\vspace{1cm}
\caption{
Invariant ratio $A_2$ vs $(\beta_{12} - \beta_{12,c}^{\rm eff}) N^\phi$
for three different pairs of $\beta_1$ and $\beta_2$. 
}
\label{Fig:A2scal}
\end{figure}

\begin{figure}
\centerline{\epsfig{file=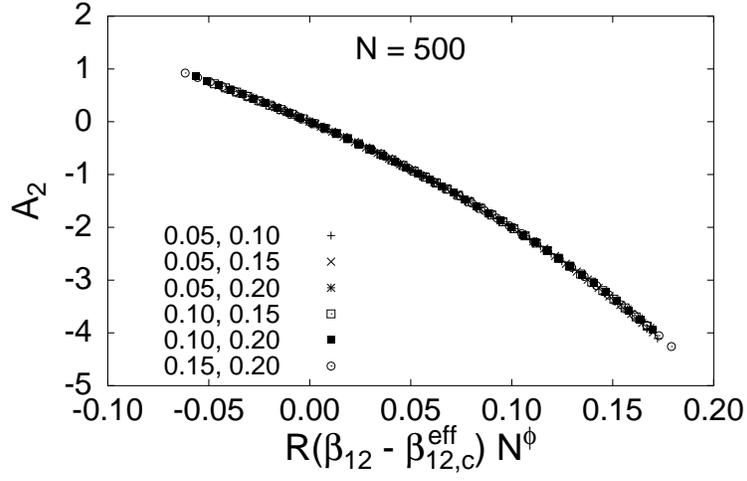,angle=-90,width=10truecm}}
\vspace{1cm}
\caption{
Invariant ratio $A_2$ vs $R (\beta_{12} - \beta_{12,c}^{\rm eff}) N^\phi$
for $N=500$. We report results corresponding to six different pairs of 
$\beta_1$ and $\beta_2$. 
}
\label{Fig:A2resc}
\end{figure}

In order to obtain a better qualitative understanding of the corrections
we have also performed additional simulations. The results are reported in 
Fig.~\ref{Fig2}. The qualitative behavior is very similar in all cases and 
almost independent of $\beta_1$ and $\beta_2$. In particular, corrections
appear to vanish in all cases for $0.05 \lesssim \beta_{12} \lesssim 0.10$
and to increase strongly for $\beta_{12} \gtrsim 0.20$. 
Morever, $\beta_{12,c}$ is always close to $\beta_\theta$, and shows a 
tiny dependence on $\beta_1$ and $\beta_2$.

The discussion presented here addressed the behavior of multicomponent 
solutions, but it should be noted that the results are also relevant 
for copolymers in which chemically different polymers
are linked together.\cite{WR-foot} Also in this case scaling corrections
with exponent $\omega_T$ are present.

Finally, we have also considered the behavior close to the ideal-mixing 
point, where there is no effective interaction between the 
two chemically different polymer species. Our numerical results 
are in very good agreement with the
renormalization-group predictions of Ref.~\CITE{SLK-91}.

\begin{table}[t]
\begin{center}
\begin{tabular}{llllcr}
type & $N_{\rm min}$ & 
\multicolumn{1}{c}{$\omega_T$} & 
\multicolumn{1}{c}{$A_2^*$} & $\chi^2$ & DOF \\
\hline
fit 1 & 100 & $0.424 \pm 0.002 \pm 0.004 $  & 
                          & 15.2 & 20 \\ 
fit 1 & 250 & $0.418 \pm 0.004 \pm 0.006 $  & 
                          & 12.4 & 17 \\  
fit 1 & 500 & $0.415 \pm 0.007 \pm 0.008 $  & 
                          & 11.8 & 14 \\  
fit 2 & 100 & $0.426 \pm 0.015 $  & $5.498 \pm 0.011 $
                          & 32.2 & 26 \\ 
fit 2 & 250 & $0.421 \pm 0.030 $  & $5.495 \pm 0.018 $
                          & 32.1 & 22 \\ 
fit 3 & 100 & $0.426 \pm 0.015 $            & $5.499 \pm 0.011 $
                          & 32.1 & 26 \\ 
fit 3 & 250 & $0.423 \pm 0.029 $            & $5.496 \pm 0.017 $
                          & 32.0 & 22 \\ 
\end{tabular}
\end{center}
\caption{Results of the fits. When two errors are quoted, the first
one is the statistical error, while the second one is the systematic
error. DOF is the the number of degrees of freedom of the fit. Fit 3
provides an estimate of $\Delta_T = \nu\omega_T$. 
We compute $\omega_T = \Delta_T/\nu$ by using $\nu = 0.5876\pm 0.0001$. 
}
\label{Table}
\end{table}

\begin{thebibliography}{199}
\bibitem{footnote}
A precise estimate of $\omega - \omega_T$ can be obtained by writing it 
as $\omega/2 - \zeta$. Using the estimate of Ref.~\CITE{BN-97} for 
$\omega$ and the estimate of $\zeta$ reported in 
Sec.~\ref{sec2}, we find $\omega - \omega_T = 0.447\pm 0.025$.
\end{thebibliography}
\end{document}